\documentclass[aps,prl,twocolumn,superscriptaddress,showpacs,preprintnumbers,amsmath,amssymb]{revtex4}
\usepackage{graphicx}
\usepackage{dcolumn}
\usepackage{bm}
\usepackage{pstricks}
\usepackage{pst-node}

\newcommand{\svalue}{-0.61}
\newcommand{\sstaterr}{\pm0.10}
\newcommand{\ssysterr}{\pm0.04}
\newcommand{\avalue}{+0.55}
\newcommand{\astaterr}{\pm0.08}
\newcommand{\asysterr}{\pm0.05}

\newcommand{\pipi}{\pi^+\pi^-}
\newcommand{\kpi}{K^+\pi^-}
\newcommand{\apipi}{{\cal A}_{\pi\pi}}
\newcommand{\spipi}{{\cal S}_{\pi\pi}}
\newcommand{\akpi}{{\cal A}_{K\pi}}
\newcommand{\qq}{q\overline{q}}
\newcommand{\de}{\Delta E}
\newcommand{\mbc}{M_{\rm bc}}
\newcommand{\dt}{\Delta t}
\newcommand{\dmd}{\Delta m_d}
\newcommand{\taub}{\tau_{B^0}}

\usepackage{epsfig}

\usepackage{relsize}
\def\babar{\mbox{\slshape B\kern-0.1em{\smaller A}\kern-0.1em
    B\kern-0.1em{\smaller A\kern-0.2em R}}}

\begin{document}



\title{\quad\\[0.5cm] \boldmath 
{Observation of Direct {\boldmath $CP$}-Violation in
$B^0\to\pipi$ Decays 
and Model-Independent Constraints on $\phi_2$}
}

\date{\today}

\affiliation{Budker Institute of Nuclear Physics, Novosibirsk}
\affiliation{Chiba University, Chiba}
\affiliation{University of Cincinnati, Cincinnati, Ohio 45221}
\affiliation{Department of Physics, Fu Jen Catholic University, Taipei}
\affiliation{University of Hawaii, Honolulu, Hawaii 96822}
\affiliation{Hanyang University, Seoul}
\affiliation{High Energy Accelerator Research Organization (KEK), Tsukuba}
\affiliation{Hiroshima Institute of Technology, Hiroshima}
\affiliation{University of Illinois at Urbana-Champaign, Urbana, Illinois 61801}
\affiliation{Institute of High Energy Physics, Chinese Academy of Sciences, Beijing}
\affiliation{Institute of High Energy Physics, Vienna}
\affiliation{Institute for Theoretical and Experimental Physics, Moscow}
\affiliation{J. Stefan Institute, Ljubljana}
\affiliation{Korea University, Seoul}
\affiliation{Kyungpook National University, Taegu}
\affiliation{Swiss Federal Institute of Technology of Lausanne, EPFL, Lausanne}
\affiliation{University of Ljubljana, Ljubljana}
\affiliation{University of Maribor, Maribor}
\affiliation{University of Melbourne, Victoria}
\affiliation{Nagoya University, Nagoya}
\affiliation{Nara Women's University, Nara}
\affiliation{National Central University, Chung-li}
\affiliation{National United University, Miao Li}
\affiliation{Department of Physics, National Taiwan University, Taipei}
\affiliation{H. Niewodniczanski Institute of Nuclear Physics, Krakow}
\affiliation{Nippon Dental University, Niigata}
\affiliation{Niigata University, Niigata}
\affiliation{University of Nova Gorica, Nova Gorica}
\affiliation{Osaka City University, Osaka}
\affiliation{Osaka University, Osaka}
\affiliation{Panjab University, Chandigarh}
\affiliation{Peking University, Beijing}
\affiliation{Princeton University, Princeton, New Jersey 08544}
\affiliation{RIKEN BNL Research Center, Upton, New York 11973}
\affiliation{University of Science and Technology of China, Hefei}
\affiliation{Seoul National University, Seoul}
\affiliation{Shinshu University, Nagano}
\affiliation{Sungkyunkwan University, Suwon}
\affiliation{University of Sydney, Sydney NSW}
\affiliation{Tata Institute of Fundamental Research, Bombay}
\affiliation{Toho University, Funabashi}
\affiliation{Tohoku Gakuin University, Tagajo}
\affiliation{Tohoku University, Sendai}
\affiliation{Department of Physics, University of Tokyo, Tokyo}
\affiliation{Tokyo Institute of Technology, Tokyo}
\affiliation{Tokyo Metropolitan University, Tokyo}
\affiliation{Tokyo University of Agriculture and Technology, Tokyo}
\affiliation{Virginia Polytechnic Institute and State University, Blacksburg, Virginia 24061}
\affiliation{Yonsei University, Seoul}
  \author{H.~Ishino}\affiliation{Tokyo Institute of Technology, Tokyo} 
  \author{K.~Abe}\affiliation{High Energy Accelerator Research Organization (KEK), Tsukuba} 
  \author{K.~Abe}\affiliation{Tohoku Gakuin University, Tagajo} 
  \author{I.~Adachi}\affiliation{High Energy Accelerator Research Organization (KEK), Tsukuba} 
  \author{H.~Aihara}\affiliation{Department of Physics, University of Tokyo, Tokyo} 
  \author{D.~Anipko}\affiliation{Budker Institute of Nuclear Physics, Novosibirsk} 
 \author{K.~Arinstein}\affiliation{Budker Institute of Nuclear Physics, Novosibirsk} 
  \author{T.~Aushev}\affiliation{Swiss Federal Institute of Technology of Lausanne, EPFL, Lausanne}\affiliation{Institute for Theoretical and Experimental Physics, Moscow} 
  \author{A.~M.~Bakich}\affiliation{University of Sydney, Sydney NSW} 
  \author{E.~Barberio}\affiliation{University of Melbourne, Victoria} 
% ...
\collaboration{The Belle Collaboration}

\begin{abstract}
We report a new measurement of the 
time-dependent $CP$-violating parameters
in $B^0\to \pi^+\pi^-$ decays with
$535\times10^6$~$B\overline{B}$ pairs collected 
with the Belle detector at the KEKB asymmetric-energy $e^+e^-$
collider operating at the $\Upsilon(4S)$ resonance.
We find $1464\pm65$ 
$B^0\to\pi^+\pi^-$ events and 
measure the $CP$-violating parameters
$\spipi  = \svalue\sstaterr ({\rm stat})
                              \ssysterr({\rm syst})$ and
$\apipi = \avalue\astaterr ({\rm stat})
                              \asysterr ({\rm syst})$. 
We observe large direct $CP$-violation with a significance 
greater than 5 standard deviations for any $\spipi$
value.
Using isospin relations, we 
measure the Cabibbo-Kobayashi-Maskawa quark-mixing matrix  
angle $\phi_2 = (97\pm11)^{\circ}$ 
for the solution consistent with the standard model
and exclude the range 
$11^{\circ}<\phi_2<79^{\circ}$ at the 95\% confidence level.

\end{abstract}


\maketitle


In the standard model (SM) framework,
$CP$ violation is attributed to an irreducible complex phase
in the Cabibbo-Kobayashi-Maskawa (CKM) weak-interaction 
quark-mixing matrix~\cite{ckm}.
In the decay chain of $\Upsilon(4S) \to B^0\overline{B}{}^0$,
one $B^0$ decays into $\pipi$ at time $t_{\pi\pi}$,
while the other decays at time $t_{\rm tag}$ into a flavor 
specific state $f_{\rm tag}$.
The time-dependent $CP$ violation~\cite{Sanda} is given as
\begin{eqnarray}
{\cal P}_{\pi\pi}^q(\Delta t) 
& = & \frac{e^{-|\Delta t|/\tau_{B^0}}}{4\tau_{B^0}}
[1 + q \cdot \{ {\cal S}_{\pi\pi} \sin(\Delta m_d \Delta t) \nonumber \\
 & &         + {\cal A}_{\pi\pi} \cos(\Delta m_d \Delta t ) \} ], 
\label{eq:signal-pdf}
\end{eqnarray}
where $\dt=t_{\pi\pi}-t_{\rm tag}$,
$\tau_{B^0}$ is the $B^0$ lifetime, $\dmd$ is the mass
difference between the two $B$ mass eigenstates~\cite{PDG2006}
and $q=+1$ $(-1)$ when $f_{\rm tag} = B^0(\overline{B}{}^0)$.
$\spipi$ and $\apipi$ are the mixing-induced
and direct $CP$-violating parameters, respectively.

The $CP$-violating parameters have been measured by the
Belle~\cite{pipi-belle} and BaBar~\cite{pipi-babar} collaborations.
Both experiments obtained consistent results for $\spipi$.
In contrast, BaBar measured an $\apipi$ value consistent with zero, 
while Belle found evidence for large direct $CP$ violation
with a significance of four standard deviations ($\sigma$)
using a data sample containing $275\times 10^6$ $B\overline{B}$ pairs.
Here we report a new measurement with 
a large data sample ($535\times 10^6$ $B\overline{B}$ pairs),
and improvements to the analysis method
that increase its sensitivity.
We confirm our earlier results and observe direct $CP$
violation in $B^0\to\pipi$~\cite{CC} decays
at the $5.5\sigma$ level;
the disagreement with the BaBar $\apipi$ measurement~\cite{pipi-babar}
remains.
This result rules out or strongly constrains
superweak models~\cite{super-weak}, extensions of the SM in which
all $CP$ violation occurs through $\Delta B=2$ processes.

One of the CKM angles, $\phi_2$~\cite{alpha}, can be measured using 
$\spipi=\sqrt{1-\apipi^2}\sin(2\phi_2 + \kappa)$, where $\kappa$ is
determined using isospin relations~\cite{isospin}.
This angle has been measured using not only $B\to\pi\pi$ decays
but also $B\to \rho\pi$ and 
$\rho\rho$ decays~\cite{rhopi-rhorho};
all the measurements give consistent results,
and the combined $\phi_2$ value, together with measurements
of other CKM angles and sides, is consistent with the 
unitarity of the matrix~\cite{ckmfit, utfit}. 
We combine our $\spipi$ and $\apipi$ measurements
with the world average (W.A.) values of other quantities
to obtain a new constraint from $B\to\pi\pi$ on $\phi_2$.
Multiple solutions are found;
for the solution consistent with other CKM measurements 
in the context of the SM,
the constraint is more restrictive than
those obtained from other $B$ decay modes.

The data sample used in this analysis was collected with the Belle
detector~\cite{belle-detector} 
at the KEKB $e^+e^-$ asymmetric-energy (3.5 on 8 GeV) 
collider~\cite{KEKB} operating at the $\Upsilon(4S)$ resonance
produced with a Lorentz boost factor of
$\beta\gamma = 0.425$ nearly along the electron beam direction ($z$ axis).
Since the two $B$ mesons are produced approximately at rest in the 
$\Upsilon(4S)$ center-of-mass system (CMS),
the decay time difference $\Delta t$ is determined from the distance
between the two $B$ meson decay vertices along the 
$z$-direction ($\Delta z$):
$\Delta t \cong \Delta z / c\beta\gamma$.
In the Belle detector, a silicon vertex detector 
and a 50-layer central drift chamber (CDC) are
used for charged particle tracking,
and an array of aerogel threshold Cherenkov counters 
as well as the $dE/dx$ measurements in the CDC provide
the particle identification (PID) 
information to distinguish charged pions and kaons.
The devices are placed inside a superconducting solenoid coil
providing a 1.5~T magnetic field. 

We employ the event selection of Ref.~\cite{pipi-belle}
except for the PID requirement, which is removed.
This increases the signal detection efficiency by 23~\%.
The PID information is instead used in a likelihood fit in this analysis,
improving the measurement errors for
the $CP$-violating parameters by about 10~\%
compared with the previous analysis. 
We reconstruct $B^0 \to \pi^+\pi^-$ candidates using oppositely
charged track pairs. 
We select $B$ meson candidates using the energy difference 
$\de \equiv E_B^*-E_{\rm beam}^*$
and the beam energy constrained mass 
$\mbc\equiv\sqrt{(E_{\rm beam}^*)^2-(p_B^*)^2}$,
where $E_{\rm beam}^*$ is the CMS beam-energy, and 
$E_B^*$ and $p_B^*$ are the CMS energy and momentum of the $B$ candidate.
We define the signal box as
$5.271$~GeV/$c^2 < \mbc < 5.287$~GeV/$c^2$ and
$|\de|<0.064$~GeV.

The standard Belle algorithm in Ref.~\cite{TaggingNIM} identifies the flavor
of $f_{\rm tag}$ using properties of its decay products,
and provides two variables: $q$ defined in Eq.~(\ref{eq:signal-pdf})
and $r$.
The parameter $r$ 
ranges from $r=0$ indicating no flavor
discrimination to $r=1$ indicating unambiguous flavor assignment.
The candidate events are categorized into six $r$ intervals ($l=1,6$).
The wrong tag fraction in each $l$ bin, $w_l$,
and the differences between $B^0$ and $\overline{B}{}^0$ decays,
$\Delta w_l$, are determined using
data~\cite{TaggingNIM}.

To discriminate the continuum background 
($e^+e^-\to q\overline{q}$, $q=u,d,s,c$),
we form signal ($q\overline{q}$ background) likelihood functions,
${\cal L}_{S(BG)}$, from features of the event topology,
and require $r$-dependent thresholds of 
${\cal R} = {\cal L}_S/({\cal L}_S + {\cal L}_{BG})$ for the candidates,
as the separation between signal and $\qq$ background depends on $r$.
The thresholds are determined to be
0.50, 0.45, 0.45, 0.45, 0.45, and 0.20 for each $l$ bin
by optimizing the expected sensitivity
using signal Monte Carlo (MC) events and events in the sideband region
$5.20$~GeV/$c^2<\mbc<5.26$~GeV/$c^2$ or 
$+0.1\,\mathrm{GeV}<\de<+0.5\,\mathrm{GeV}$.
We further divide the data sample into two categories having
${\cal R}$ above or below $0.85$ to take into account the
correlation between the $\de$ shape of 
$q\overline{q}$ background and ${\cal R}$.
We thus have 12 distinct bins of ${\cal R}$ {\sl vs.} $r$; 
these bins are labeled $\ell=1,6$ ($\ell=7,12$) for
the six $r$ intervals with ${\cal R}>0.85$ $({\cal R}<0.85)$.

We extract signal candidates in the global area $\mbc>5.20$~GeV/$c^2$
and $-0.3$~GeV$<\de<+0.5$~GeV
by applying the above requirements
and the vertex reconstruction algorithm used in Ref.~\cite{btos-belle}.
The selected candidates are not only $B^0\to\pipi$ signal events
but also include $B^0\to\kpi$, $q\overline{q}$ 
and three-body $B$ decay backgrounds.
We estimate the signal and background yields with an unbinned 
extended maximum likelihood fit, making use of $\de$, $\mbc$
and the kaon identification probability 
$x_{\pm}={\cal L}_{K^{\pm}}/({\cal L}_{K^{\pm}}+{\cal L}_{\pi^{\pm}})$
for the positively and negatively charged tracks
of $B^0\to\pipi$ candidates,
where ${\cal L}_{\pi^{\pm}}$ and ${\cal L}_{K^{\pm}}$ are the likelihood values 
for the pion and kaon hypotheses.

We use a sum of two bifurcated Gaussians and a single Gaussian
to model the $\de$ and $\mbc$ shapes, respectively,  
for both $B^0\to\pipi$ and $\kpi$.
The probability density functions (PDF) as a function of $x_{\pm}$ for 
the signal and $B^0\to\kpi$ decays are obtained from
a large data sample of inclusive $D^{*+}\to D^{0}\pi^{+}, 
D^{0}\to K^{-}\pi^{+}$ decays.
The yields of $B^0\to K^{+}\pi^{-}$
events are parameterized as $n_{K^{\pm}\pi^{\mp}}= n_{K\pi}(1\mp\akpi)/2$,
where $\akpi=-0.113\pm0.020$~\cite{PDG2006} is 
the direct $CP$ asymmetry 
in $B^0\to\kpi$ decays.
We fix the $\akpi$ value and float the $B^0\to\kpi$ yield
$n_{K\pi}$ in the fit.
For the signal and $B^0\to\kpi$ events, we use MC-determined event
fractions in each ${\cal R}$-$r$ bin.

The $q\overline{q}$ background shapes in $\de$ and $\mbc$ are described
by a second-order polynomial and an ARGUS function~\cite{argus}, 
respectively.
We use $qr$-dependent two-dimensional $(x_+, x_-)$ PDFs
for the $q\overline{q}$ background
to take into account the correlation between positively and negatively
charged tracks.
These PDFs are determined from the sideband events.

For the three-body $B$ decays,
we employ a smoothed two-dimensional histogram obtained from
a large MC sample for the $\de$-$\mbc$ shape.
We use the same $x_{\pm}$ PDFs as those of the signal 
and $B^0\to\kpi$ decays, 
with a $\de$-dependent kaon fraction determined from the MC sample.

\begin{figure}
\epsfysize=2.7cm  \epsfbox{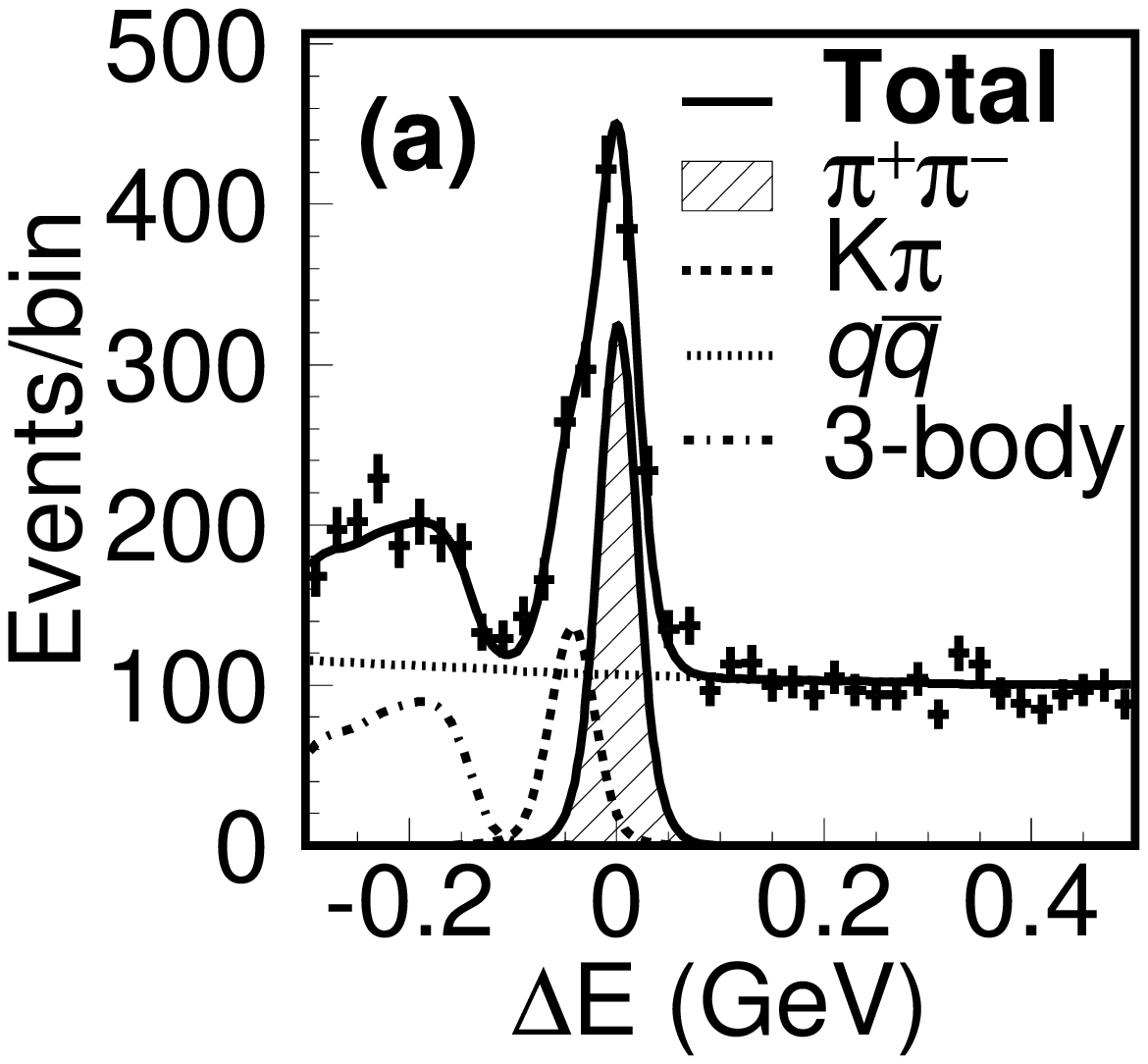}
\epsfysize=2.7cm  \epsfbox{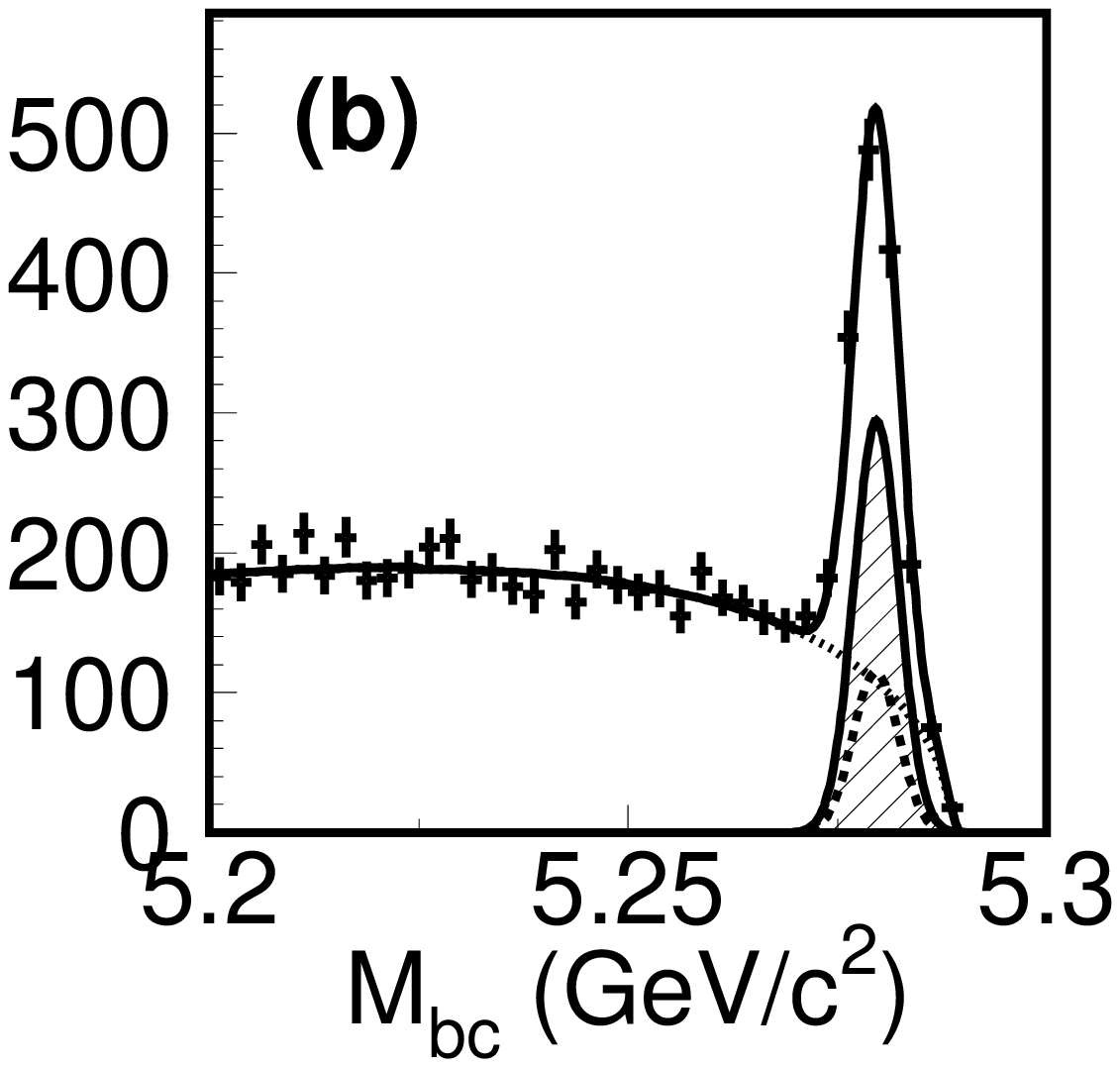}
\epsfysize=2.7cm  \epsfbox{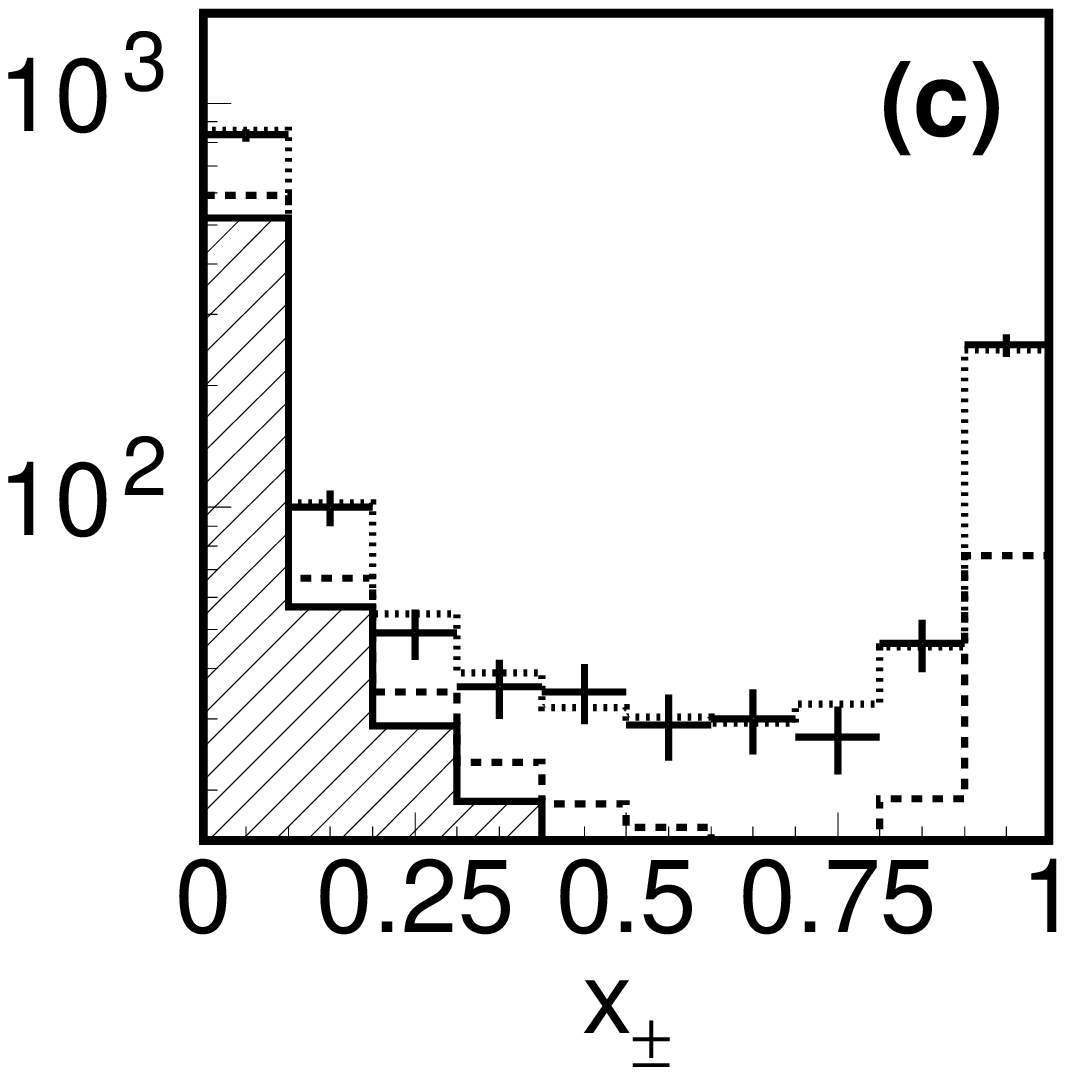}
\caption{
(a) $\de$, (b) $\mbc$ and (c) $x_{\pm}$ projection plots
of the $B^0\to\pipi$ candidates having ${\cal R}>0.85$
in the signal box of (a) $\mbc$ with $x_{\pm}<0.4$, 
(b) $\de$ with $x_{\pm}<0.4$ and (c) $\mbc$ with
$0$~GeV$< \de <0.02$~GeV.
Figure (c) is the sum of $x_+$ and $x_-$ distributions.
}
\label{fig:de-mbc}
\end{figure}

By fitting to the data in the global area we determine the yields
of the signal and background components.
Interpolating to the signal box,
we obtain $1464\pm65$ $\pipi$,
$4603\pm105$ $\kpi$ and $10764\pm33$ $q\overline{q}$ events,
where the errors are statistical only.
The contribution from three-body $B$ decays is negligible in the
signal box.
From the signal yield and the detection efficiency (53.1~\%),
we estimate the measured branching fraction to be $(5.2\pm0.2)\times 10^{-6}$,
in agreement with the W.A. value~\cite{HFAG}.
Figure~\ref{fig:de-mbc} shows the projection plots of $\de$, $\mbc$
and $x_{\pm}$ for candidate events. 

To determine $\spipi$ and $\apipi$,
we apply an unbinned maximum likelihood fit to the $\dt$ distribution
of the 16831 candidates in the signal box.
The signal distribution 
in Eq.~(\ref{eq:signal-pdf}) is modified
to incorporate the effect 
of incorrect flavor assignment, using $w_l$ and $\Delta w_l$.
This distribution is then convolved with the proper time interval
resolution function $R_{\rm sig}(\Delta t)$~\cite{vertexres}.
The final signal PDF is given by 
$P_{\pi\pi}^{q\ell}(\dt)=(1-f_{\rm ol}){\cal P}_{\pi\pi}^{q\ell} (\dt)
\otimes R_{\rm sig}\,
(\dt)+f_{\rm ol}{\cal P}_{\rm ol}(\dt)$,
where the outlier PDF ${\cal P}_{\rm ol}(\dt)$ accommodates a small
fraction $f_{\rm ol}$ of events having large $\dt$ values.
The $\dt$ distribution for $B^0\to K^{+}\pi^{-}$ is
${\cal P}_{K^{\pm}\pi^{\mp}}^{q\ell}(\dt) = (1/4\taub)e^{-|\dt|/\taub}
[1-q\Delta w_l \mp q(1-2w_l)\cos(\dmd\dt)]$;
the corresponding PDF  
$P_{K^{\pm}\pi^{\mp}}^{q\ell}(\dt)$
is constructed in the same manner as the signal PDF.
The $q\overline{q}$ background distribution 
contains prompt and finite-lifetime components;
it is convolved with a background resolution
function modeled as a sum of two Gaussians
and combined with the outlier PDF to give the $q\overline{q}$
background PDF $P_{q\bar{q}}(\dt)$.
All the parameters of $P_{q\bar{q}}(\dt)$ are determined
using sideband events.

We define a likelihood value for the $i$-th event,
which lies in the $\ell$-th bin of ${\cal R}$ {\sl vs.} $r$:
\begin{equation}
P_i = \sum_k n_k^{\ell}{\cal P}_k^{q(\ell)}(\vec{s}_i)P_k^{(q\ell)}(\dt_i).
\end{equation}
Here $n_k^{\ell}$ is the fraction of component
$k \in \{\pi^+\pi^-, K^+\pi^-, K^-\pi^+, q\bar{q} \}$
in ${\cal R}$-$r$ bin $\ell$;
${\cal P}_k^{q(\ell)}(\vec{s})$ is the event-by-event probability
for component $k$ as a function of $\vec{s}= (\de, \mbc, x_+, x_-)$; 
and $P_k^{(q\ell)}(\dt)$ is the event-by-event probability for component
$k$ and flavor tag $q$ as a function of $\dt$.
In the fit, $\spipi$ and $\apipi$ are the only free parameters
and are determined by maximizing the likelihood function
${\cal L}=\prod_i P_i$.

\begin{figure}
\epsfysize=6.0cm  \epsfbox{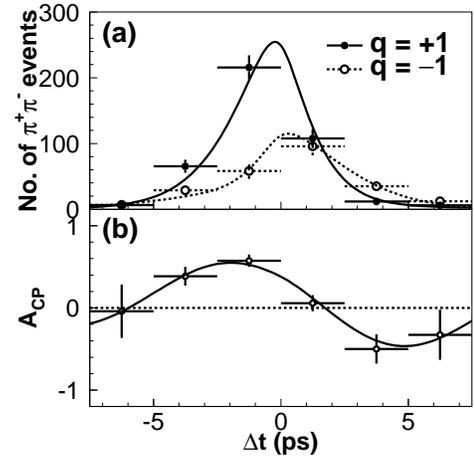}
\caption{
(a) $\dt$ distributions of $B^0\to\pipi$ signal events 
with $r>0.5$ after background subtraction for $q=+1$ (solid)
and $q=-1$ (dashed),
and (b) asymmetry ${\cal A}_{CP}$ plot.
The curves are projections of the fit result.
The difference in the heights of the $q=+1$ and $q=-1$ components
in (a) is due to direct $CP$ violation.
}
\label{fig:cp-asym}
\end{figure}

The unbinned maximum likelihood fit yields 
$\spipi=\svalue\sstaterr({\rm stat})\ssysterr({\rm syst})$
and
$\apipi=\avalue\astaterr({\rm stat})\asysterr({\rm syst})$.
The correlation between $\spipi$ and $\apipi$ is $\rho=+0.15$.
Figures~\ref{fig:cp-asym} (a) and (b) show the background subtracted $\dt$
distributions of the signal events with $r>0.5$ for $q=\pm1$ and
the asymmetry ${\cal A}_{CP}$ in each $\dt$ bin, respectively, where
${\cal A}_{CP}=(N_{+}-N_{-})/(N_{+}+N_{-})$ and $N_{+(-)}$
is the number of signal events with $q=+1$ $(-1)$
obtained by a fit in each $\dt$ bin.

The main contributions to the systematic error are
due to uncertainties in the vertex reconstruction
($\pm0.03$ for $\spipi$ and $\pm0.01$ for $\apipi$)
and event fractions 
($\pm0.01$ for $\spipi$ and $\pm0.04$ for $\apipi$);
the latter includes a conservative uncertainty for the possible 
$q\overline{q}$ background flavor asymmetry of $\pm0.02$.
We include the effect of tag side interference~\cite{TSI}
on $\spipi$ ($\pm0.01$) and $\apipi$ ($\pm0.02$).
Other sources of systematic error are the uncertainties
in the wrong tag fraction 
($\pm0.01$ for both $\spipi$ and $\apipi$), 
physics parameters ($\taub$, $\dmd$ and $\akpi$)
($<0.01$ for both $\spipi$ and $\apipi$), 
resolution function
($\pm0.02$ for both $\spipi$ and $\apipi$), 
background $\dt$ shape
($<0.01$ for both $\spipi$ and $\apipi$), 
and fit bias
($\pm0.01$ for both $\spipi$ and $\apipi$).
We add each contribution in quadrature to obtain the total
systematic error.

To validate our $CP$-violating parameter measurement,
we check the measurement of $\apipi$ using a time-integrated fit,
and obtain $\apipi=+0.56\pm0.10$, consistent with the time-dependent
fit results.
An unbinned extended maximum likelihood fit to the
$q=+1$ ($q=-1$) subset with ${\cal R}>0.85$ and $r>0.5$ 
yields $280\pm20$ ($169\pm16$) $\pi^+\pi^-$ 
signal events, in agreement with the measured $\apipi$ value taking into
account the dilution due to 
the wrong tag fractions and $B^0 \bar{B}{}^0$ mixing.
We also check the direct $CP$ asymmetry in $B^0\to\kpi$ events by
floating $\akpi$ in the time-dependent fit,  
and obtain a value consistent with the W.A.~\cite{PDG2006}
and the same $\rho$ value with the nominal fit.
The fit is applied to various data subsets: 
a subset containing events with positive (negative) $\de$ 
in which the $B^0\to\kpi$ contamination is suppressed (enriched),
where $\apipi=+0.60\pm0.11$ $(+0.51\pm0.12)$,
events with ${\cal R}>0.85$ $({\cal R}<0.85)$ where 
the $q\overline{q}$ background 
fraction is suppressed (enriched),
events with $x_{\pm}<0.4$ where the signal fraction is enhanced,
and events in one of the six $r$ bins having different wrong
tag fractions.
All fits to the subsets yield $CP$ asymmetries consistent with the
overall fit result. 
We also carry out a fit to the sideband events,
and find no sizable asymmetry.

\begin{figure}
\epsfxsize=6cm  \epsfbox{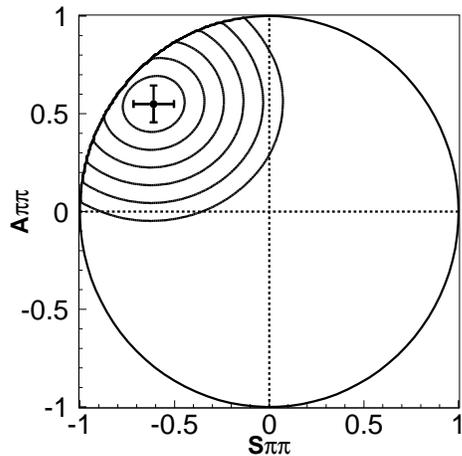}
\caption{
Confidence regions for $\spipi$ and $\apipi$.
The curves show the contour for
$1-{\rm C.L.}=3.17\times 10^{-1}$ $(1\sigma)$, 
$4.55\times 10^{-2}$ $(2\sigma)$,
$2.70\times 10^{-3}$ $(3\sigma)$, 
$6.33\times 10^{-5}$ $(4\sigma)$, 
$5.73\times 10^{-7}$ $(5\sigma)$ 
and $1.97\times 10^{-9}$ $(6\sigma)$
from inside to outside.
The point with error bars is the ${\cal S}_{\pi\pi}$
and ${\cal A}_{\pi\pi}$ measurement.
}
\label{fig:cl-level}
\end{figure}

We determine the statistical significance of our measurement
using a frequentist approach~\cite{fc}, taking into account both
statistical and systematic uncertainties.
Figure~\ref{fig:cl-level} shows the resulting two-dimensional
confidence regions in the $\spipi$ and $\apipi$ plane.
The case of no direct $CP$ violation, $\apipi=0$, is ruled out
at a confidence level (C.L.) of $1- 4 \times 10^{-8}$,
equivalent to a $5.5\sigma$ significance
for one-dimensional Gaussian errors.
We also observe mixing-induced $CP$ violation 
with a significance greater than $5.3\sigma$ for any $\apipi$ value.

\begin{figure}
\epsfxsize=6cm  \epsfbox{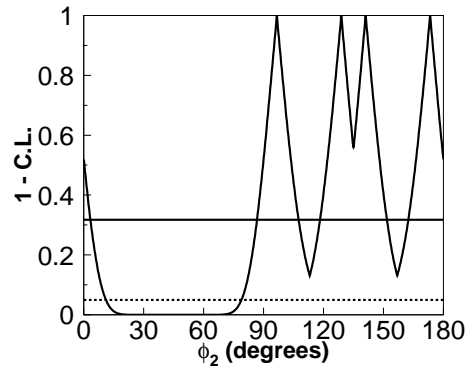}
\caption{
Difference 1-C.L. plotted for a range of $\phi_2$ values
obtained with an isospin analysis
using Belle measurements of ${\cal S}_{\pi\pi}$ 
and ${\cal A}_{\pi\pi}$, 
and the W.A. values for the direct $CP$ asymmetry 
in $B^0\to\pi^0\pi^0$ decays and branching fractions
for the $B\to\pi\pi$ modes.
The solid and dashed lines indicate C.L.=68.3\% and 95\%, respectively.
}
\label{fig:phi2}
\end{figure}

To constrain $\phi_2$, 
we use the isospin relations~\cite{isospin}
with our measured values of ${\cal S}_{\pi\pi}$
and ${\cal A}_{\pi\pi}$,
the W.A. branching ratios
of $B^0\to \pi^+\pi^-$ ($5.2\pm 0.2$ in units of $10^{-6}$), 
$\pi^0\pi^0$ ($1.31\pm 0.21$)
and $B^+\to \pi^+\pi^0$ ($5.7\pm0.4$),
and the W.A. direct $CP$-asymmetry for $B^0 \to \pi^0\pi^0$ 
($0.36\pm0.33$)~\cite{HFAG}.
The isospin relations imply that
the branching ratios and $CP$-violating parameters 
can be expressed with six parameters, one of which is $\phi_2$.
We construct a $\chi^2$ using 
the measured values and the six parameters.
We follow the statistical method of Ref.~\cite{ckmfit}.
We minimize the $\chi^2(\phi_2)$ for each $\phi_2$ value
by varying the remaining five parameters with constraints on 
the corresponding branching ratios and asymmetries
and compute the C.L. from the cumulative distribution of
the difference $\Delta \chi^2 = \chi^2(\phi_2) - \chi^2_{\rm min}$
for one degree of freedom,
where $\chi^2_{\rm min}$ is the minimum $\chi^2(\phi_2)$ value.
Figure~\ref{fig:phi2} shows the difference $1 - $C.L. plotted for
a range of $\phi_2$ values.
We find four different solutions consistent with our measurement.
For the solution consistent with the expectation 
from other CKM measurements, $(100^{+5}_{-7})^{\circ}$~\cite{ckmfit},
we find $\phi_2 = (97\pm11)^{\circ}$. 
We exclude the $\phi_2$ range 
$11^{\circ}<\phi_2<79^{\circ}$ at the 95\% confidence level.

In summary, 
using a data sample containing 535$\times 10^6 B\overline{B}$ pairs
we measure the $CP$-violating parameters in $B^0\to\pipi$ decays:
$\spipi=\svalue\sstaterr({\rm stat})\ssysterr({\rm syst})$
and
$\apipi=\avalue\astaterr({\rm stat})\asysterr({\rm syst})$.
We report the first observation of direct $CP$ violation with
$5.5\sigma$ significance.
Our results as well as the evidence for direct $CP$ violation 
in $B^0\to \kpi$
decays~\cite{kpi-directcp} rule out superweak models~\cite{super-weak}. 
The measured $\spipi$ and $\apipi$ values in this Letter 
are consistent with those reported in Ref.~\cite{pipi-belle},
and supersede Belle's earlier evidence for direct $CP$ violation.
Among the four $\phi_2$ solutions, 
the $\pm 1\sigma$ range for the $\phi_2$ solution consistent
with the SM is more restrictive 
than that from measurements of $B\to\rho\pi$ and 
$B\to\rho\rho$ decays~\cite{rhopi-rhorho}.

We thank the KEKB group for excellent operation of the
accelerator, the KEK cryogenics group for efficient solenoid
operations, and the KEK computer group and
the NII for valuable computing and Super-SINET network
support.  We acknowledge support from MEXT and JSPS (Japan);
ARC and DEST (Australia); NSFC and KIP of CAS (China); 
DST (India); MOEHRD, KOSEF and KRF (Korea); 
KBN (Poland); MIST (Russia); ARRS (Slovenia); SNSF (Switzerland); 
NSC and MOE (Taiwan); and DOE (USA).


\clearpage
\newpage


\begin{thebibliography}{999}
\bibitem{CC}
Throughout this paper, 
the inclusion of the charge conjugate mode decay is implied
unless otherwise stated.
\bibitem{alpha} Another naming convention, 
$\alpha$ ($= \phi_2$) is also used in the literature.
\end{thebibliography}
\end{document}